\documentclass[letterpaper]{article}

\usepackage{natbib}
 \usepackage{hyperref}
\usepackage{url}
\usepackage{breakurl}
\usepackage{geometry}
\usepackage{graphicx}

\usepackage{epsfig,color}

\usepackage[T1]{fontenc}
\usepackage[sc,osf]{mathpazo}

\usepackage{sectsty}
\sectionfont{\rmfamily\mdseries\Large}
\subsectionfont{\rmfamily\mdseries\itshape\large}

\renewenvironment{itemize}{
  \begin{list}{}{
    \setlength{\leftmargin}{1.5em}
  }
}{
  \end{list}
}

{\end{list}}

\begin{document}


\noindent {\huge Phylogenetic analysis of gene expression}


\vspace{0.25in}

\noindent \emph{Abstract:}

Phylogenetic analyses of gene expression have great potential for addressing a wide range of questions. These analyses will, for example, identify genes that have evolutionary shifts in expression that are correlated with evolutionary changes in morphological, physiological, and developmental characters of interest. This will provide entirely new opportunities to identify genes related to particular phenotypes. There are, however, three key challenges that must be addressed for such studies to realize their potential. First, gene expression data must be measured from multiple species, some of which may be field collected, and parameterized in such a way that they can be compared across species. Second, it will be necessary to develop phylogenetic comparative methods suitable for large multidimensional datasets. In most phylogenetic comparative studies to date, the number $n$ of independent observations (independent contrasts) has been greater than the number $p$ of variables (characters). The behavior of comparative methods for these classic $n>p$ problems are now well understood under a wide variety of conditions. In gene expression studies, and studies based on other high-throughput tools, the number $n$ of samples is dwarfed by the number $p$ of variables. The estimated covariance matrices will be singular, complicating their analysis and interpretation, and prone to spurious results. Third, new approaches are needed to investigate the expression of the many genes whose phylogenies are not congruent with species phylogenies due to gene loss, gene duplication, and incomplete lineage sorting. Here we outline general project design considerations for phylogenetic analyses of gene expression, and suggest solutions to these three categories of challenges. These topics are relevant to high-throughput phenotypic data well beyond gene expression.

\vspace{0.25in}

\noindent \emph{Authors:}

Casey W. Dunn, Assistant Professor, Department of Ecology and Evolutionary Biology, Brown University. casey\_dunn@brown.edu
\vspace{0.25in}

Xi Luo, Assistant Professor, Department of Biostatistics and Center for Statistical Sciences, Brown University. xi\_luo\_1@brown.edu
\vspace{0.25in}

Zhijin Wu, Associate Professor, Public Health-Biostatistics, Brown University. zhijin\_wu@brown.edu

\pagebreak

\section{Introduction}

RNA-seq now enables inexpensive gene expression studies in a broad diversity of species \citep{tHoen:2008hn, Siebert:2011gv}. To date, most such studies have focused on making comparisons \emph{within} species. They have, for example, examined differences in gene expression between different experimental conditions, disease status, cell types, tissue types, or genetic backgrounds. 

Many questions of great interest, however, will require phylogenetic comparative analyses of gene expression \emph{across} species. The phylogenetic analysis of gene expression, both by itself and in combination with other types of phenotypic data, will generate biological insight in a variety of ways:

\begin{itemize}
\item - Sets of genes with correlated evolutionary changes in expression will reveal shared function, shared expression regulation mechanisms, or both.
\item - An investigator often knows of only one or two genes involved in a particular biological process, but wants to find others. These known genes will be used as ``bait'' to identify other genes with correlated evolutionary changes in expression. These are strong candidates for also being involved in the biological process of interest.
\item - Significant evolutionary changes in the covariance structure of expression data may indicate evolutionary changes in gene regulation, function, or both.
\item - Gene expression will be analyzed in combination with other data, such as physiological or morphological measurements, to identify genes with evolutionary changes in expression that are correlation with evolutionary changes in specific biological processes of interest.
\item - Phylogenetic analyses of gene expression will test alternative hypotheses about how selection acts on expression. It will allow for rigorous tests of the ortholog conjecture \citep{Nehrt:2011ka}, as well as specific models of gene function evolution such as DDC \citep{Force:1999tg}. This will also provide a much better understanding of neutral expression evolution. 
\item - Comparative analyses of gene expression across species will be of great use even when an investigator is concerned with only a single species. This is because RNA-seq and other high-throughput tools are so powerful that they often detect hundreds or even thousands of genes with significant differential expression between treatments.  It is still very difficult, though, to identify which differential expression is biologically meaningful. The most informative and cost effective way to understand expression data from any particular species may be to collect similar data from closely related species, and analyze them in a combined phylogenetic analysis.
\end{itemize}

It is not statistically valid to simply analyze the correlation of expression across species. This is because observations of any trait (including expression) made across multiple species are not independent, because some species are more closely related to each other than to others. If these evolutionary relationships between species are not taken into account, one can be severely misled by the strong similarity between closely related species and the many differences that are expected to arise by chance between distantly related species. The seminal paper by \cite{Felsenstein:1985ua} demonstrated this problem and introduced an ingenious solution, phylogenetic independent contrasts. Based on the structure of the species phylogeny, it transforms the original dependent observations into a series of statistically independent contrasts. There is one independent contrasts for each of the internal nodes on the phylogeny. Phylogenetic independent contrasts can then be analyzed to assess the correlation of the measured variables through the course of evolution. \cite{Felsenstein:2008cl} has since expanded upon the original independent contrasts to account for variation within species. There is now a rich set of methods for examining the evolution of quantitative characters, most of which have been developed to examine morphological and ecological data but have not yet been applied to functional genomic data.

To apply these tools to evolutionary analyses of gene expression, we must overcome three specific analysis challenges. First, we must measure and parameterize expression data so that they can be compared across species. Second, we must confront the statistical challenges that arise when the number of variables under consideration far outnumbers the observations available. Third, new comparative methods must be developed that can accommodate gene-specific data, such as expression, when gene phylogenies are not congruent with species phylogenies.

There has long been interest in comparing gene expression across species. Using microarray data, \cite{Rifkin:2003cj} compared differential expression between two developmental time points across six lineages of \emph{Drosophila}. More recently, \cite{Brawand:2011du} investigated the evolution of expression in six organs across ten species of amniotes (one bird and nine mammals). Both studies included phylogenies that were inferred from the expression data. Neither of these studies, however, mapped the expression data onto phylogenies and perform independent contrasts or other comparative analyses.

There have been other studies and reviews that have considered various aspects the evolution of gene expression. \cite{Romero:2012hu} made a survey of the potential for comparative studies of expression to reveal the evolution of regulatory mechanisms. \cite{Gilad:2006fv} reviewed the evidence for different types of selection on expression, and concluded that stabilizing selection dominates in most cases. \cite{Zheng:2011cw} reviewed the current understanding of regulatory variation between species. \cite{HodginsDavis:2009im} highlighted the importance of taking environmental effects into account, which can greatly complicate and even mislead expression studies that consider multiple species. These broad perspectives on the evolution of expression highlight the great potential for phylogenetic comparative analyses of gene expression, once key technical challenges are addressed.

\section{Project design} 
\begin{figure}
\includegraphics[width=0.9\textwidth]{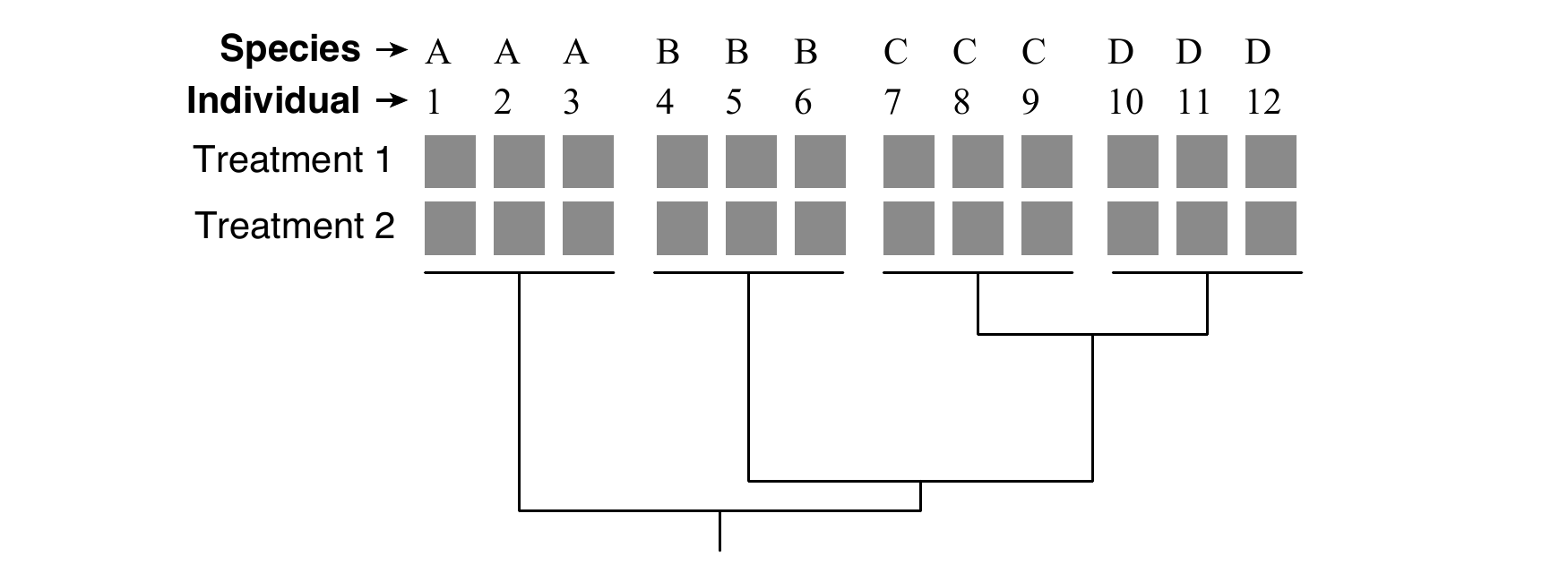}
\caption{\label{fig:tree} A typical project design for phylogenetic analysis of gene expression. Tree depicts the phylogeny of the species, and each of the grey boxes represents a sample. A read count (i.e., an expression measurement) is available for each gene for each sample. In this example, there are twelve individuals sampled across four species. In each individual, expression data are available for two treatments (e.g., tissue types). In some projects, data for different treatments may come from different individuals.}
\end{figure}

Before describing the challenges that are specific to phylogenetic analyses of gene expression data, it is first necessary to address project design (Figure \ref{fig:tree}). This is the single most important aspect of an expression study--if the data are not collected in such a way that they can answer the questions at hand then they will be useless. There are many general considerations to take into account when designing an RNA-seq study, regardless of whether the data are to be analyzed in a phylogenetic perspective or not. Most are the exact same issues that have been identified and addressed in more than a decade of micro-array studies, though unfortunately many quantitative RNA-seq studies are recapitulating the problems that were identified in early micro-array studies. There are also additional project design issues that are specific to phylogenetic analysis. 

In an RNA-seq study, mRNA is isolated from each \emph{sample} and shotgun sequenced. The resulting reads are then mapped to gene reference sequences, and the numbers of reads that map to each gene are counted to give quantitative indications of expression levels. These counts are then normalized across samples to account for differences in sequencing efforts \citep{Hansen:2012kb, Robinson:2010dd}. There is therefore one normalized read count for each gene for each sample. 

The reference sequences to which the reads are mapped to derive coutns can be derived from fully annotated genomes, or from transcriptome assemblies if genome sequences are note available. Transcriptome assemblies can be based on the same data that are used to quantify assembly, usually by assembling across all samples prior to mapping each sample individually. In most cases, especially as the number of samples grows, it is more cost effective to use long-read sequencing to assemble a high quality reference and short-read sequencing across samples to quantify expression \citep{Siebert:2011gv}.

Expression data from a single sample isn't very interesting. It can't be related to biological variation, and it isn't even possible to compare expression across genes for technical reasons described below. Gene expression studies therefore consider differences in expression across multiple samples. In general, the goal of an expression study is to identify which genes have a greater difference in expression between \emph{treatments} than would be expected by chance. Multiple samples are therefore collected across multiple treatments. ``Treatments'' is used generically here for any biological variation, whether it is experimentally induced or not. Treatments could represent control and pharmacological treatments, different tissue types, different cell types, different developmental stages, different environmental conditions, or any of a number of other differences.  

In order to have a sense of how much variation is expected by chance, it is critical to collect replicate samples for each treatment. Replication isn't just an expensive technical nuisance, it is the investigator's friend. By revealing how much variation there is when measuring expression across samples within a treatment, it provides a much better understanding of how to interpret variation across samples.

Replicate samples are usually collected across \emph{individuals} (or, sometimes, pools of individuals, e.g. clutches of embryos that are each from a single spawning event). In some cases, each sample comes from a different individual. This is necessary when the treatment affects the individual as a whole, as would be the case for a drug treatment or environmental stress. In other cases, it is possible to collect samples for different treatments from the same individual, as when comparing expression across tissue types. It is usually desirable to collect samples for different treatments from he same individual when possible, as this provides the opportunity to consider individual effects as well as treatment effects.

The aspects of project design considered above apply to all expression studies. In a phylogenetic study of gene expression, there is one more component to experimental design-- \emph{species} (Figure \ref{fig:tree}). Each individual belongs to a particular species, and there can (and should be) multiple individuals per species. This makes it possible to look at treatment, individual, and species effects on expression levels. Ideally, samples are available for each treatment for each species and, if possible, samples for different treatments are taken from the same individual (Figure \ref{fig:tree}). 

When collecting expression data from wild-caught specimens, as is likely to be the case in many studies that consider multiple species, there are many potential sources of variation that could complicate the interpretation of expression variation between and within species \citep{HodginsDavis:2009im}. This is because wild-collected specimens usually have unknown environmental histories and genetic backgrounds. Expression from two samples from two individuals may differ because one ate last week and one ate an hour ago, not because the samples are drawn from two treatments. In some cases different species may live in different habitats, and expression differences between these species may be due to differences in environment rather than due to evolutionary changes in expression. The specifics of minimizing variation due to these extraneous factors will differ from study to study, but are very important to consider. Common-garden approaches and thorough replication are two general strategies that should be used when possible.

\section{Challenge I: Measuring and parameterizing RNA-seq data so that it can be compared across species}

\subsection{The problem}

The normalized read counts produced by a quantitative RNA-seq study are not direct measurements of expression. These counts are proportional to expression, but they are also impacted by other effects that can differ across genes and species. This is because the probability of sequencing a read for a gene is impacted by both the sequence of a gene \citep{Hansen:2010if, Hansen:2012kb} and its length. These effects can be modeled with unknown species and gene specific counting efficiency coefficients. For gene $g$ in species $s$, treatment $t$, let $k_{gs}$ denote the gene and species specific counting efficiency, the expectation of gene count $C_{gst}$ is proportional to both $k_{gs}$ and the gene expression level $E_{gst}$.

\begin{equation}\label{eq:expression}
E[C_{gst}]=k_{gs}E_{gst}
\end{equation}

Since the counting efficiency $k_{gs}$ is inconsistent across genes and species, the direct comparison using $C_{gst}$ can be misleading. This is because count differences may simply be due to differences in $k_{gs}$. Imagine the trivial example where a given gene is twice as long in one species as another, but the expression level (i.e., the number of transcripts per cell) is the same. The number of counts for this gene will differ by a factor of two across species, even though there has been no evolutionary change in expression level. If not accounted for, this could severely mislead comparative analyses. The same types of impacts can be realized if a given gene has a sequence that is sequenced more efficiently than the different sequence of the same gene in another species. 

Addressing $k_{gs}$ is especially important when reference sequences are incomplete, as reads that map outside the reference will not be counted towards a gene. If the reference sequence for a gene is complete in one species but not another, then the number of reads that map to the incomplete reference sequence will be an underestimate relative to the number for the species with with the complete transcript prediction for the gene. As phylogenetic comparative studies of expression are likely to include species with reference sequences of varying quality, the ability to minimize the impacts of these differences on the evolutionary interpretation of expression is critical. If the reference sequences for all species are based on well annotated genomes, it may be the case that it is not as critical to account for $k_{gs}$. A recent study of expression across mammals and a bird, for example, directly compared length and abundance normalized counts across species \citep{Brawand:2011du}. The references for all species were derived from relatively well annotated genome projects and, although this approach did not take into account differences in sequence composition or other unknown factors that could contribute to $k_{gs}$, there were still many robust patterns across species.

Below we outline two distinct approaches to addressing the technical challenges imposed by $k_{gs}$. Each has its own advantages and drawbacks, and the approach chosen for a particular project will depend on details of study design and further evaluations of these methods. 

\subsection{A solution: Evolutionary analyses of expression ratios}

In the first analysis approach, $k_{gs}$ is canceled out within species before any comparisons are made across species. Rather than compare counts across species, the investigator compares ratios of expected counts across species. Given Equation \ref{eq:expression} above, consider the comparison of the ratio of gene expression in tissue type $1$ to tissue type $2$:

\begin{equation}
\frac{E[C_{gs1}]}{E[C_{gs2}]} = \frac{k_{gs}E_{gs1}}{k_{gs}E_{gs2}} = \frac{E_{gs1}}{E_{gs2}}
\end{equation}

Because $k_{gs}$ is in both the numerator and the denominator, the ratio of counts is determined exclusively by the ratio of expression. These ratios can then be compared across species and genes without the confounding effects of $k_{gs}$. This ratio is the Fold Change (FC), which is already widely used in visualizations of differential expression data. FC has been compared across species in an analysis of the evolution of differential expression in \emph{Drosophila} based on microarray data \cite{Rifkin:2003cj}.

The advantage of this approach is that it greatly simplifies downstream analyses since $k_{gs}$ is removed prior to any phylogenetic comparative analyses. The drawback is that, by transforming all observations to ratios, downstream tests of significance cannot take into consideration the absolute magnitude of counts. The observed variance across replicates could, however, still be used to assess significance. If the treatment samples are taken from the same individuals, as in Figure \ref{fig:tree}, the independent contrasts can be computed from the transformed data according to \cite{Felsenstein:2008cl}, which accommodates multiple individuals per species.

\subsection{A solution: Consider measurements from different treatments as different characters}

In the second approach, gene-specific counting efficiencies are not addressed prior to phylogenetic comparative analyses. Rather than consider the ratio of counts for the same gene across treatments, the counts for each gene for each treatment are initially considered as if they are different characters. If, for example, there are measurements for 5,000 genes in two treatments, these measurements are treated as if they are 10,000 different characters. The phylogenetic independent contrasts of these characters are then calculated as usual, and the covariance matrix estimated from these contrasts. Particular cells in the resulting covariance matrix will correspond to covariances between the same treatment for the same gene (i.e., the variances), different treatments for the same gene, the same treatment for different genes, or different treatments for different genes. The covariance matrix can then be decomposed into these various subsets depending on the questions at hand.

There are several advantages to this approach. It preserves the magnitudes of the normalized count data, which improves the ability to assess the significance of differences in expression. It is also a very general framework that allows for more complex experimental designs. The primary disadvantage is that it greatly increases the dimensionality of the problem, which in turn exacerbates the challenges described in the next section.

\section{Challenge II: Large number of genes and a small number of species}

\subsection{The problem}
In most studies that make use of phylogenetic independent contrasts, the number $n$ of phylogenetic independent contrasts has been far greater than the number $p$ of variables. The behavior of these classic $n>p$ evolutionary problems are now well understood under a wide variety of conditions. In phylogenetic analyses of expression based on high-throughput expression data, though, the number $n$ of independent contrasts (i.e., the number of species minus one) is dwarfed by the number $p$ of variables (the number of genes considered, if differential expression ratios are plotted onto the trees, or the number of treatments times the number of genes, if the counts for each gene for each treatment are considered as separate characters). These new $n<<p$ analyses raise a variety of challenges.

\begin{figure}
\includegraphics[width=0.9\textwidth]{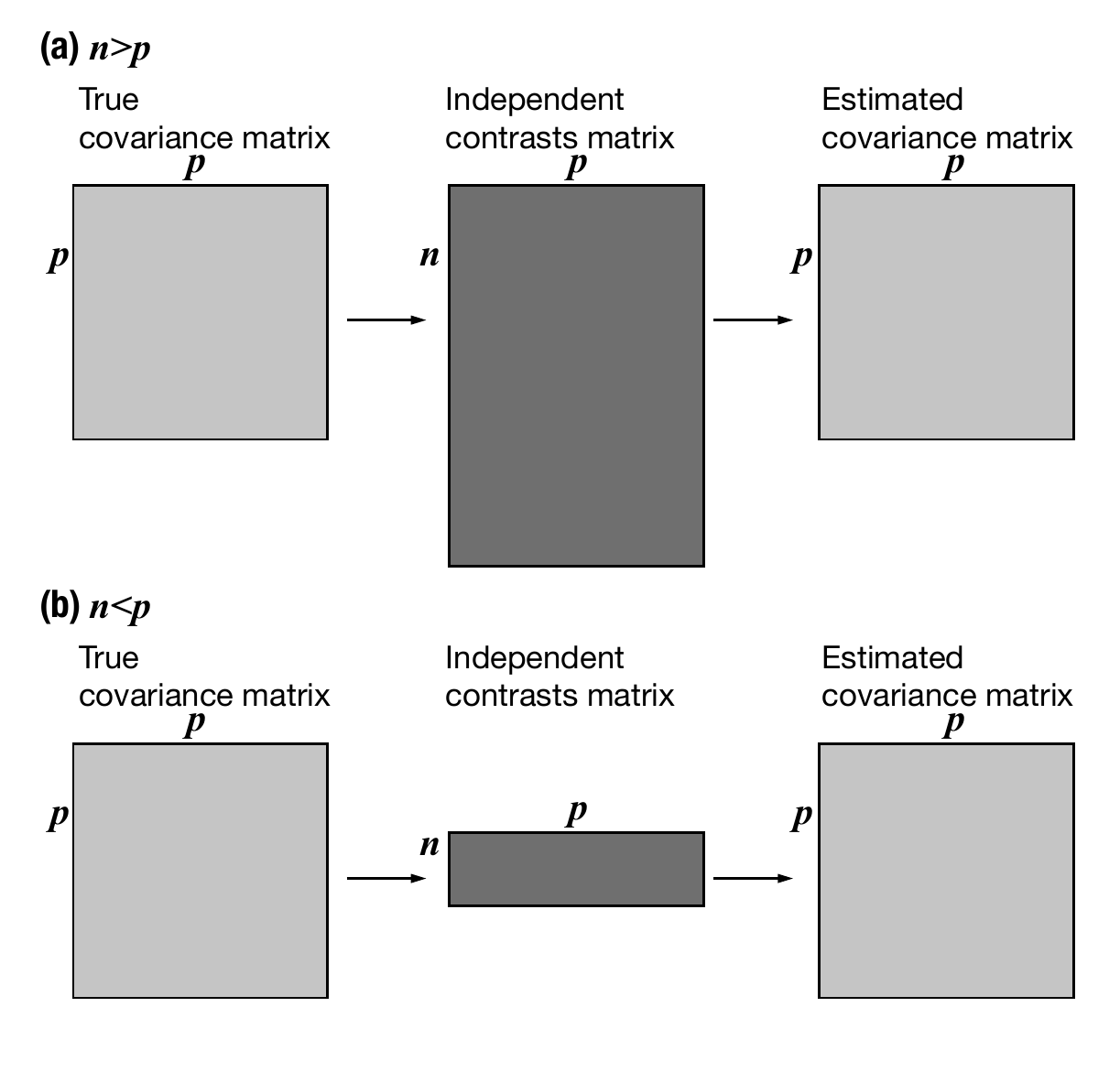}
\caption{\label{fig:space} An illustration of relative matrix dimensions in comparative analyses. $n$ is the number of observations (independent contrasts) and $p$ is the number of variables (characters) measured in each observation. (a) $n>p$, so the independent contrasts matrix is larger than the covariance matrices and it is possible to uniquely estimate the true covariance matrix. This is the case for most comparative studies to date, which consider many more species than variables. (b) $n<p$, so the independent contrasts matrix is smaller than the covariance matrices and it is not possible to uniquely estimate the true covariance matrix. This is the situation for phylogenetic comparative analyses of RNA-seq expression data, as well as other high throughput phenotype data.}

\end{figure}

The crux of the problem is that when $n<p$, the information provided by the contrasts is not sufficient to uniquely construct the covariance matrix from observed data (Figure \ref{fig:space}). The covariance matrix is always of size $p \times p$. This applies to both the true, but unknown, covariance matrix, as well as the covariance matrix that is estimated from the observed data. The matrix of observations (i.e., independent contrasts), however, has size $n \times p$. When $n>p$, this matrix is larger than the covariance matrix and can contain enough information to uniquely infer each element of the covariance matrix. When $n<p$, this matrix is smaller than the covariance matrix and the elements of the inferred covariance matrix cannot be uniquely determined.

In essence, when $n<<p$ the true covariance matrix is being squeezed through the much smaller data matrix and then expanded back out to the observed covariance matrix (Figure \ref{fig:space}). It is analogous to compressing and then expanding a digital photograph. If the original photograph is a thousand by a thousand pixels, and you compress it to a format that is a thousand elements by ten elements, there is no way to uniquely reconstruct each pixel of all possible original photographs.

More formally, when $n<p$ the estimated covariance matrix will be singular. Singular matrices are not invertible, which means that many basic linear algebra manipulations can't be done exactly. There are several important implications of this. Some common statistical procedures cannot be performed on singular covariance matrices. It also means that the number of principal components that could be derived from the estimated covariance matrix will be determined by the number of independent observations, not the number of variables. A covariance matrix of size $p  \times p$ could have up to $p$ principal components (one corresponding to each eigen vector). In the singular matrices described above, however, only $n$ of these will correspond to non-zero eigen values.

\subsection{A solution}
To address the $n<p$ challenge in the context of phylogenetic analyses of gene expression data, we propose to adapt a recently developed and powerful statistical framework, matrix regularization. Compared to alternative approaches, such as principal component analysis, regularization has advantages in computational speed, interpretability, robustness, and good power in small sample sizes \citep{Hastie:2009wp}. These regularization approaches try to match the observed data with a set of most likely patterns arising from a model. They remain data-driven because the target pattern is general enough that only significant correlations will be retained by the regularization methods. 

\begin{figure}
\includegraphics[width=0.9\textwidth]{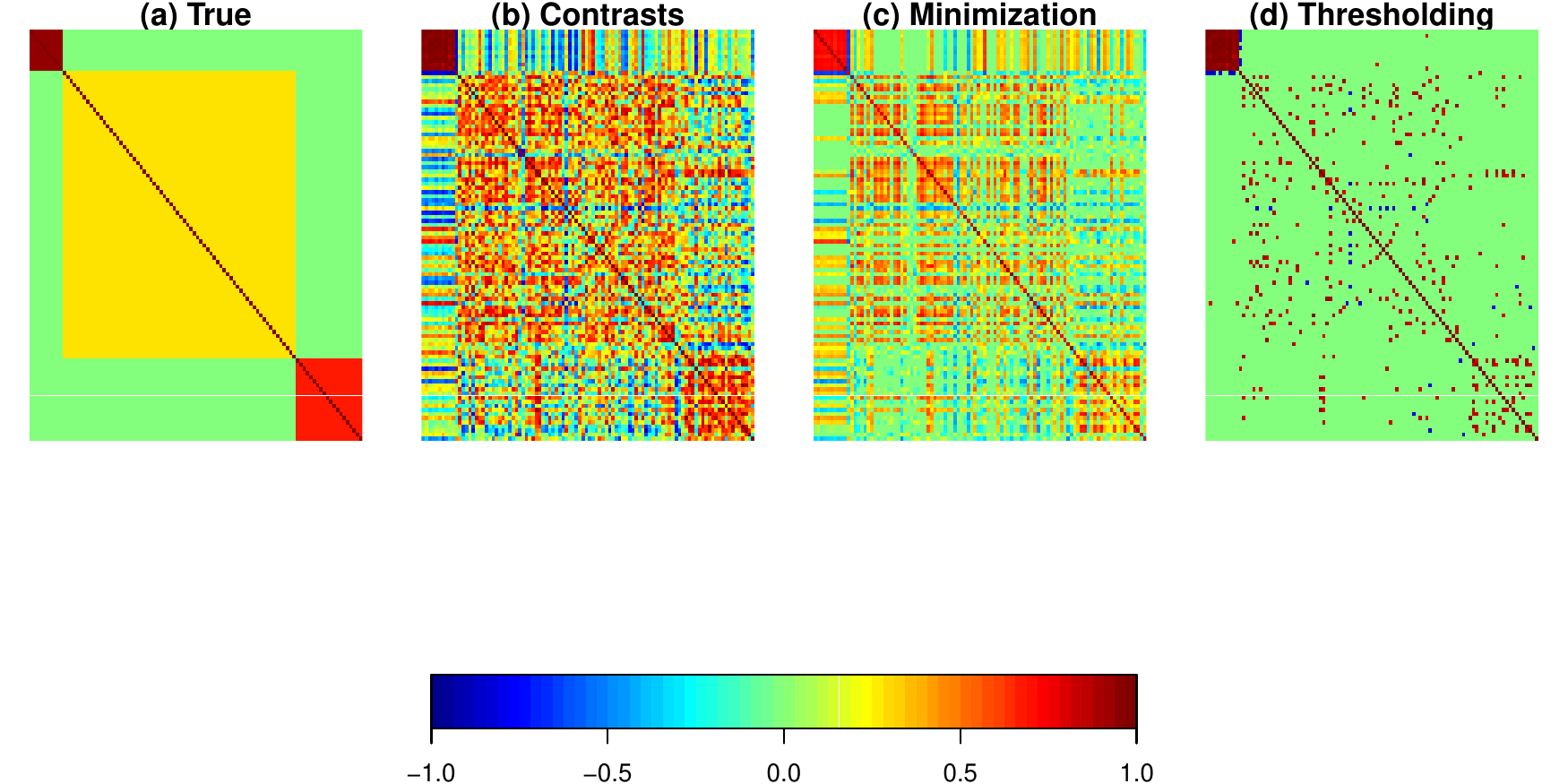}
\caption{\label{fig:simu} Simulation analyses of correlation matrix reconstruction. The evolution of expression of 100 genes was simulated on an 8 taxon phylogeny. The true covariance matrix is block-diagonal (a). The other three matrices (b-d) show the results of alternative approaches to reconstructing the correlation matrix. The correlation matrix inferred directly from the independent contrasts has spurious high and low correlations for many genes that do not have covariance (b). Regularization of matrix b reduces the number of these false positives (c-d). Convex minimization \citep[c, ][]{Luo:2011tw} is less conservative than thresholding \citep[d, ][]{Bickel:2008kl}.}
\end{figure}

Figure \ref{fig:simu} presents the result of simulation analyses, comparing the true covariance matrix (Figure \ref{fig:simu}a) to inferred covariance matrices (Figure \ref{fig:simu}b-d). We considered one hundred variables (which could each be the ratio of expression of one hundred genes in two treatments) on an eight taxon tree. The source code for these analyses is available at \url{https://bitbucket.org/caseywdunn/sicb2013}.

It is clear that the covariance matrix derived directly from the independent contrasts (Figure \ref{fig:simu}b) is extremely noisy in comparison to the true covariance matrix that was used to simulate the data. Strong covariance (both negative and positive) is spuriously detected between genes that have no true covariance. Though strong covariance is recovered for genes that do indeed have strong covariance (see the upper left of Figure \ref{fig:simu}b), some genes with moderate covariance are found to have nearly zero covariance. This noise is expected since the $n \times p$ independent contrast matrix does not have enough information to uniquely estimate the $p \times p$ covariance matrix (Figure \ref{fig:space}). 

Here we consider two methods of regularization, convex minimization \citep{Luo:2011tw} and thresholding \citep{Bickel:2008kl}. Thresholding is more conservative than convex minimization, with fewer false positives but many more false negatives. In convex minimization, covariance matrix is decomposed into two components. The first component is attributed to the effects of unmeasured factors (e.g., environmental effects and regulation pathways) and the second component is attributed to strong pairwise gene expression correlations after accounting for the unmeasured factor effects . The statistical model behind this decomposition is related to other popular models, including principal component analysis and surrogate variable analysis \citep{Leek:2007kn}, but generalizes to exploit the covariance structures in terms of both eigenvalues/eigenvectors and matrix entries. In thresholding, only strong (both positive and negative) pairwise gene correlations are retained, which parallels the second component of convex minimization but without accounting for unmeasured factors. Both regularization methods require input parameters on the regularization strength trading off false positives and false negatives, and we here only consider the theoretical choices as illustrations. Additional statistical research is needed for designing robust, interpretable and data-driven ways for choosing the regularization strength in this context. 

These results indicate that regularization can at least in part overcome some of the challenges that arise in the phylogenetic comparative analysis of high-dimensional functional genomic data. The exact approach that is taken for each study will depend on the goals. If the investigator would like to identify a small number of genes with high effect and avoid false positives, then a conservative regularization approach such as thresholding \citep{Bickel:2008kl} would be appropriate. If the goal is to identify the greatest number of genes that covary with a particular phenotypic character, then a less conservative approach such as convex minimization \citep{Luo:2011tw} would be appropriate. If no regularizations are applied, great care must be taken in interpreting covariances, even when they are inferred to be quite strong. 

\section{Challenge III: Accommodating gene duplication and loss}
\subsection{The problem}
In the analyses above, we assumed that each species had the exact same set of genes. In reality, genes are duplicated and lost through the course of evolution. This leads to the expansion, refinement, and even complete loss of gene families in different species through time. As a result, genes have phylogenetic histories that are not always the same as the phylogenetic histories of the species under consideration (Figure \ref{fig:paralogs}). 

Previous high-throughput studies of gene expression across species have focused on the subset of genes that have only strict orthologs \citep{Rifkin:2003cj,Brawand:2011du}. This approach greatly simplifies analyses, but discards a large fraction of the data and precludes the investigation of many phenomena of broad interest, such as the evolution of gene expression following gene duplication.

\subsection{A solution}

\begin{figure}
\includegraphics[width=0.9\textwidth]{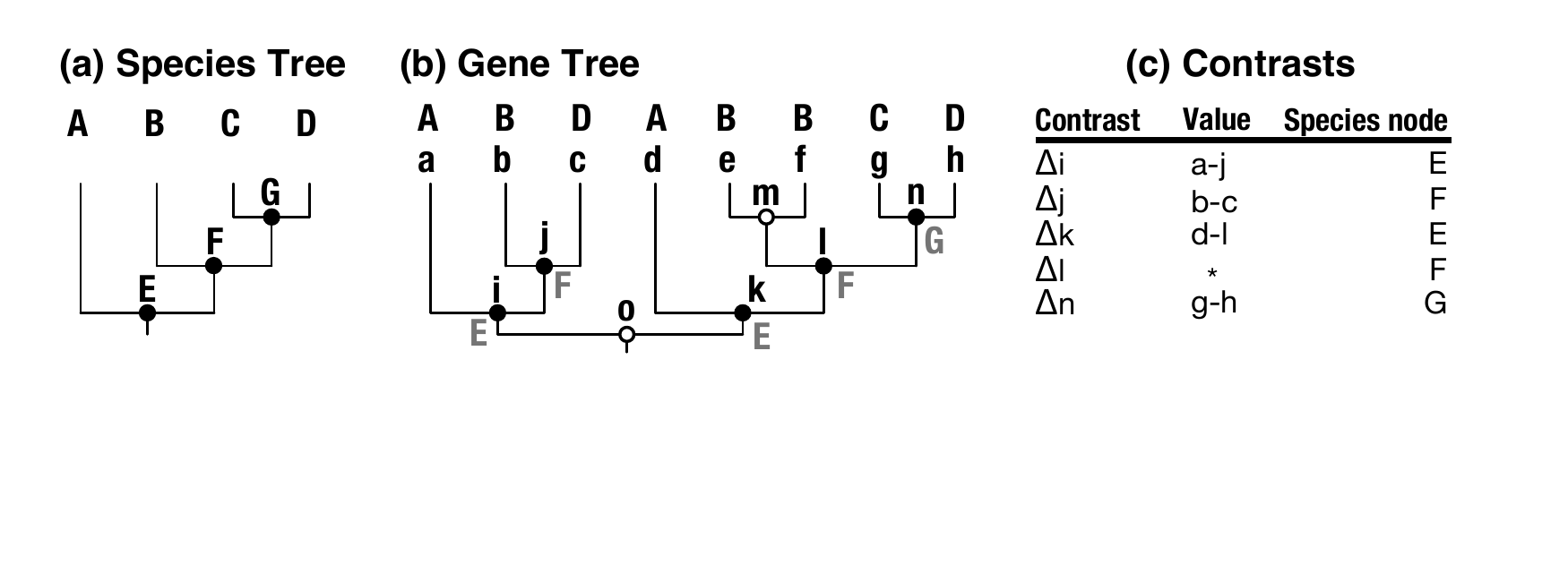}
\caption{\label{fig:paralogs} Calculating contrasts associated with speciation events on a gene tree that includes paralogs. (a) The species phylogeny. Nodes representing species and speciation events are labeled with capital letters. (b) The gene phylogeny, which includes paralogs and orthologs. Black nodes indicate speciation events, white nodes indicate gene duplications. The tips and internal nodes of the gene phylogeny are labeled with lower case letters. The species that each gene is drawn from is indicated with a capital letter, and internal speciation nodes are labeled with a capital letter for the corresponding node in the species tree. (c) Table of contrasts that correspond to speciation events.  There are five contrasts in the gene tree that together represent all three speciation events in the species tree. These contrasts can then be combined across gene trees to form an independent contrasts matrix, where the values each gene tree correspond to a column and the rows correspond to species nodes (when there are multiple values from a given species node in a gene tree, they could be averaged before being added to the contrasts table). Calculating the contrast for node l (indicated by *) is complicated by the fact that one of the descendent nodes is a duplication. Such contrasts could be skipped, or accommodated with more complicated calculations.}
\end{figure}

Several methods that reconcile gene phylogenies to species phylogenies have recently been developed \citep{Akerborg:2009dj,Arvestad:2009ky,Sennblad:2009gy}. These tools label each node in a gene phylogeny as either a speciation event or gene duplication event. By definition, the speciation nodes of the gene trees each correspond to particular nodes in the species tree. This means that a common ontology can be used for these speciation nodes across all gene trees and the species tree. Each duplication node may by unique to a particular gene tree, as when a new copy of a particular gene arises by tandem duplication. Such unique duplication events must be named individually. It may also be that there are duplication events that are shared by multiple genes, such as whole genome duplication events. If there as additional external information that allows these types of duplications to be labeled in each gene trees, then a common name can be used for the shared duplication node across all gene trees. In any particular gene tree, the same speciation node or shared duplication node may be present multiple times.

Once each node in each gene tree is labeled as a speciation, shared duplication, or unique duplication node, it is possible to proceed with phylogenetic independent contrasts across all gene trees. The calculation of contrasts associated with speciation events is the most straightforward (Figure \ref{fig:paralogs}). For each internal node in the species tree, identify the two descendent nodes as in a typical contrast analysis (for example, the descendants of F are B and G in Figure \ref{fig:paralogs}a). Then, find the corresponding internal node (in Figure \ref{fig:paralogs}b, nodes j and l correspond to speciation event F) and descendent nodes (in Figure \ref{fig:paralogs}b, nodes b, e, and f correspond to node B in the species tree, and node n corresponds to speciation node G) in each gene tree, and calculate the contrasts based on the difference in expression values as reconstructed on the gene tree. One complication is that the number of nodes for a given speciation event will not be consistent across gene trees. There are various ways to address this, the most simplistic would be to take the average contrast value across the different nodes of a gene tree that correspond to the same speciation event.

In this way, the investigator builds up the contrasts that correspond to each internal node on the species tree across all gene trees. This results in a $n \times p$ contrast matrix, where $n$ is the number of internal nodes on the species tree and $p$ is the number of gene trees. The $p \times p$ covariance matrix can then be estimated from this matrix of contrasts. A similar approach could be taken for calculating contrasts across shared duplication events. Covariance  cannot be computed across unique duplication events since they are gene-specific. Variances (after normalizing by branch length) could be compared across the different categories of nodes to see if significantly larger changes are realized for one category of nodes relative to the others.

Depending on the objectives of a study, it may be desirable to consider only a subset of the contrasts that are made on a gene tree. This is because a given contrast on the species tree may also span one or more duplication events on the gene tree. This is the case for contrast l in Figures \ref{fig:paralogs}b-c. This means that the observed differences may be due to duplication effects as well as evolution between species. To avoid these potentially confounding complications, the investigator could consider only the contrasts where the descendent node are connected to the internal node by unbroken branches that do not include duplication nodes.

This general approach could be expanded to accommodate other sources of incongruence between gene trees and species trees, such as incomplete lineage sorting.

\section{Conclusion}

Though there are multiple challenges that must be addressed to enable phylogenetic comparative analyses of high-throughput gene expression data, they are not insurmountable. Some solutions, such as the use of count ratios rather than counts, can be immediately implemented with off-the-shelf tools. Others will require further methods refinement and implementation with new tools.

Several of the approaches presented here are applicable to any high dimensional quantitative phenotypic data, not just gene expression. They will therefore be useful for interpreting other categories of functional genomic and proteomic data, as well as the high-throughput approaches just now being applied to morphology, development, and physiology. It is likely that the greatest biological insight will come from combining these different types of data into single analyses, allowing for the examination of functional genomic and other phenotypic data in a single evolutionary framework.

\section{Acknowledgements} 
We thank Joe Felsenstein for insightful conversations that helped to frame and approach the problem. This work was supported by the NSF Waterman Award.

\section{Author contributions} 
CWD conceived of and initiated the study, and wrote most of the manuscript. XL and ZW developed and applied the regularization approaches. All three authors wrote the code together for the simulations and analyses.

\section{Additional files} 
The R code that produced the simulations and generated Figure \ref{fig:simu} can be found at \url{https://bitbucket.org/caseywdunn/sicb2013}.

\pagebreak

\bibliographystyle{natbib}

\bibliography{Dunn-etal}

\end{document}